\documentclass[%
 reprint,
superscriptaddress,
nofootinbib,
amsmath,amssymb,
aps,
pra,
]{revtex4-2}

\usepackage{graphicx}
\usepackage{dcolumn}
\usepackage{bm}
\usepackage{braket}
\usepackage[dvipsnames]{xcolor}
\usepackage{color}
\usepackage{amsmath}
\usepackage{ulem}

\begin{document}

\preprint{APS/123-QED}

\title{Unit-density SU(3) Fermi-Hubbard Model with Spin Flavor Imbalance}

\author{ Zewen Zhang}
\affiliation{Department of Physics and Astronomy, Rice University, Houston, Texas 77005, USA}
\author{ Qinyuan Zheng}
\thanks{Current address: Yale University, New Haven, Connecticut}
\affiliation{Department of Physics and Astronomy, Rice University, Houston, Texas 77005, USA}
\author{Eduardo Ibarra-García-Padilla}
\thanks{Current address: Sandia National Laboratories}
\affiliation{Department of Physics and Astronomy, Rice University, Houston, Texas 77005, USA}
\affiliation{Department of Physics and Astronomy, San Jos\'e State University, San Jos\'e, California 95192, USA}
\affiliation{Department of Physics and Astronomy, University of California, Davis, California 95616, USA}
\author{Richard T. Scalettar}
\affiliation{Department of Physics and Astronomy, University of California, Davis, California 95616, USA}
\author{ Kaden R. A. Hazzard}
\affiliation{Department of Physics and Astronomy, Rice University, Houston, Texas 77005, USA}
\affiliation{ Smalley-Curl Institute, Rice University, Houston, Texas 77005, USA}

\date{\today}

\begin{abstract}
The advent of ultracold alkaline-earth atoms in optical lattices has established a platform for investigating correlated quantum matter with SU($N$) symmetry, offering highly tunable model parameters that allow experiments to access phenomena that are unavailable in conventional materials. 
Understanding the ground-state physics of SU($N$) Fermi-Hubbard models away from the Heisenberg limit and from the spin-flavor balanced setting is important, as examining the flavor imbalance reveals new physics in Fermi-Hubbard models and shows how SU($N$) phases react to practical experimental imperfections in optical lattices.
In this study, mean-field phase diagrams are presented for the unit-density SU(3) Fermi-Hubbard model at two sets of flavor densities, 
$\left(\tfrac{1}{3}-\delta,\tfrac{1}{3}+\delta,\tfrac{1}{3}\right)$ and $\left(\tfrac{1}{4}-\delta,\tfrac{1}{4}+\delta,\tfrac{1}{2}\right)$, with the flavor imbalance introduced as $\delta$. 
Novel phases are identified at moderate interaction strengths for both densities and their robustness is investigated in the presence of flavor imbalance. Furthermore, we provide microscopic explanations of the phases found and their stability.  
Analysis of thermal ensembles of random mean-field solutions indicate that, at temperatures accessible in state-of-the-art cold atom experiments, some spin orders are hard for conventional scattering or local observable measurements to detect, but can be more accessible with quantum gas microscopy in optical lattice experiments.
This work also shows that nesting and Mottness, intertwined in the usual SU(2) Hubbard model in stark contrast to generic materials, can be tuned in the SU(3) model and play distinct roles. The resulting phase diagrams not only deepen our understanding of SU($N$) Fermi-Hubbard models but also inform future experimental search for new phases.
\end{abstract}

\maketitle


\section{Introduction}
The SU($N$) Fermi-Hubbard model up to densities around unit-filling has been realized in alkaline-earth-atom (AEA) optical lattices~\cite{taie20126,scazza2014observation,ozawa2018antiferromagnetic,tusi2022flavour,ferraretto2023enhancement}. Various Mott insulators and magnetic correlations have been observed~\cite{taie20126,scazza2014observation,ozawa2018antiferromagnetic,tusi2022flavour,ferraretto2023enhancement,taie2022observation,hofrichter2016direct}. 
Ongoing advances in quantum gas microscopy~\cite{bakr2009quantum,sherson2010single,taie2010realization,hofrichter2016direct,okuno2020schemes,buob2024strontium,su2024fast} (QGM) are bringing us closer to unraveling the physics of these models, 
but most of the phase diagrams of these models remain unexplored.
One key goal is to elucidate ground-state properties in SU($N$) systems, particularly with the strong quantum fluctuations associated with their enhanced symmetry. The extra spin degree of freedom in SU($N$) systems also provides us opportunity to address Hubbard physics in a more flexible way than standard SU(2) Fermi-Hubbard models.

At unit-density, the SU($N$) Fermi-Hubbard model exhibits richer physics than the  SU(2) version~\cite{affleck1988large,hermele2009mott,chen:synthetic_2016,lee2018filling, hafez2020interaction,ibarra2021universal,perez2021phase,basak2023generalized,ibarra2024many,pasqualetti2024equation,botzung2024numerical}.
In the spin-balanced SU(2) model at unit-density, the system is always in the N\'eel antiferromagnetic state, even at arbitrarily small $U$, due to the Fermi surface nesting.
In contrast, SU($N$) models near unit-density generally lack nesting unless one flavor population is specifically brought to half-filling, 
and consequently a metallic phase survives at finite interaction strength and one or more magnetic phases are expected. 
Focusing on the simplest case, $N=3$ in a two-dimensional (2D) square lattice, studies (via exact diagonalization~\cite{toth2010three}, DMFT~\cite{PhysRevA.80.051602,sotnikov2015critical}, 
DMRG~\cite{bauer2012three,schlomer2024subdimensional}, and Monte Carlo simulations~\cite{motegi2023thermal,ibarra2023metal,feng2023metal}) 
have identified a metallic phase in the weak-interaction regime and a three-sublattice magnetically ordered state in the Heisenberg limit. Research~\cite{sotnikov2015critical,ibarra2023metal,feng2023metal} has also suggested at least one intermediate phase between these two limits.

Flavor imbalance gives an extra dimension to elucidate the properties of SU(3) phases, including understanding the competition among them and justifying their robustness against the limitation in ongoing cold-atom experiments~\cite{ottenstein2008collisional,bonnes2012adiabatic,muller2021state}---not only in AEA optical lattices but also in potential multi-component alkali atom~\cite{ottenstein2008collisional,mongkolkiattichai2025quantum} and ultracold molecule~\cite{mukherjee2024n,mukherjee2024n1} experiments.
Flavor imbalance has been widely explored in the context of cold atoms, such as in Fermi gas~\cite{revelle20161d,del2018selective,sundar:spin-imbalanced_2020,he2024thermodynamics} or in Hubbard models slightly away from half-filling~\cite{honerkamp2004ultracold}. 
In the attractive SU(2) Fermi gas, the FFLO phase induced by spin imbalance is reported in theory~\cite{mathy2011trimers,sun2013pair,sundar2020spin,inotani2021radial,lydzba2020unconventional} and experiments~\cite{revelle20161d}.
In SU(2) Hubbard models, earlier investigations demonstrated that imbalanced hopping amplitudes (or effective masses)~\cite{dao2012mott,liu2015quantum} 
and spin polarization~\cite{brown2017spin} can produce diverse phase diagrams. 
Also, in SU($N$) Hubbard models, previous works indicate interesting physics when the Hamiltonian's SU($N$) symmetry is explicitly broken, such as flavor-selective phases induced by spin-dependent on-site interactions~\cite{tusi2022flavour}.
By tuning flavor imbalance to break the SU($N$) symmetry in Fermi-Hubbard systems with $N \ge 3$, 
we can gain valuable insights into these models and chart new directions for cold atom experiments.
On the other hand, practical issues, such as imperfect pumping in optical lattice experiments, can inadvertently introduce flavor imbalance, 
making it critical to assess how these imbalances affect SU($N$) experiment results.

Finally, adding flavor imbalance to the SU($N$) Hubbard models allows one to separate the effects of nesting and Mottness (insulation driven by interactions at integer filling). 
For the half-filled SU(2) Hubbard model on a square lattice, the nesting enhances Fermi surface instabilities and favors spin-density wave (SDW) order at the nesting vector $(\pi,\pi)$, while strong interactions suppress double occupancy, further solidifying the insulating behavior. For this SU(2) system, nesting and Mottness necessarily coincide (a rare coincidence not shared by generic materials or models) for both spin flavors, but these are separate for SU($N$) systems.  
While the effects of nesting and Mottness are separated in certain SU($N$) systems~\cite{wang2019slater,wang2022transition}, adding spin imbalance to these systems imparts further freedom in studying Mottness versus nesting, as individual components can be nested separately, while retaining the Mott condition. Understanding these systems could further illuminate the interplay of these mechanisms in general Fermi-Hubbard models.

In this paper, we calculate the flavor-imbalanced phase diagram of the SU(3) Fermi Hubbard model at the unit-density regime based on the Hartree mean-field approximation. We concentrate on two families of flavor densities:  $\left(\tfrac{1}{3}-\delta,\tfrac{1}{3}+\delta,\tfrac{1}{3}\right)$ and $\left(\tfrac{1}{4}-\delta,\tfrac{1}{4}+\delta,\tfrac{1}{2}\right)$, with $\delta$ the portion of imbalance. In the former, for $\delta$ not too large, none of the flavors are nested, while in the latter, for any $\delta$, one flavor remains nested. 

At moderate interaction strengths and for the flavor densities 
$\bigl(\tfrac{1}{3}-\delta,\tfrac{1}{3}+\delta,\tfrac{1}{3}\bigr)$, 
we identify three ordered phases for $\delta=0$ with quite different robustness against flavor imbalance, and show that this can be understood within a local perturbative framework. 
When both the interaction strength $U$ and imbalance ratio $\delta$ are sufficiently large, 
we additionally find $(\pi,\pi)$ structure factor peaks, signifying the same N\'eel ordering pattern as seen in the standard SU(2) model, 
despite the absence of nesting in any spin flavor.

Two ground states are found for the 
$\left(\tfrac{1}{4}-\delta,\tfrac{1}{4}+\delta,\tfrac{1}{2}\right)$ 
flavor densities. The phase diagram with flavor imbalance reveals how the SU(3) model transitions toward the SU(2) limit 
as two of the flavors move closer to nesting: the calculations uncover regions with phase separation that bridge the SU(3) and SU(2) phases.

We also examine the thermal ensemble of mean-field states at temperature currently achievable in leading AEA optical lattice experiments. With the survey of random states thermal-averaged according to their energies, we observe that certain ordered phases are more readily detected at such temperatures via 
scattering or measurements of local observables, whereas others require more advanced techniques. Achieving lower temperatures or employing QGM may be necessary to observe these more elusive phases.

Although the mean-field calculations are an uncontrolled approximation, they are expected to broadly offer a good guide to the potential structures that can occur in the model, and moreover they offer valuable guiding wavefunctions for some quantum Monte Carlo methods that are free from the sign problem, such as the constrained path quantum Monte Carlo algorithm (CP-QMC), which can offer high-accuracy calculations~\cite{zhang1997constrained,qin2016coupling}, extending the recent application on the SU(3) flavor-balanced case~\cite{feng2023metal}.

Our paper is structured as follows. Sec.~\ref{sec:model} introduces the SU($N$) Fermi-Hubbard models and the Hartree approximation we implement. Sec. \ref{sec:phase_D} discusses phase diagrams of the flavor-imbalanced SU(3) Hubbard model, both at zero temperature and finite temperature.
Sec.~\ref{sec:conclusion} concludes.

\section{Model and Method}
\label{sec:model}
The SU($N$) Fermi-Hubbard model on a 2D square lattice is
\begin{equation}
    \label{Hamiltonian}
    H = -t\sum_{\langle i,j\rangle,\sigma}
    c^\dagger_{i,\sigma}c^{\phantom \dagger}_{j,\sigma} +
    \frac{U}{2}\sum_{i,\sigma_1\neq\sigma_2}
    n_{i,\sigma_1} n_{i,\sigma_2},
\end{equation}
where $c_{i,\sigma}$ is the fermionic annihilation operator of spin flavor $\sigma$ at site $i$, $n_{i,\sigma}=c_{i,\sigma}^\dagger c^{\phantom \dagger}_{i,\sigma}$ is the corresponding number operator, $\langle i,j\rangle$ indicates pairs of nearest-neighbor sites, $t$ is the tunneling rate, and $U$ is the interaction strength. In AEA optical lattice experiments, the spin-independent $U$ arises because the nuclear spin is decoupled from the electronic structure in the ground state~\cite{wu2003exact,gorshkov2010two,cazalilla2014ultracold}. This Hamiltonian conserves individual spin flavor populations.

The Hartree approximation expands the particle number operator around a mean-field value, \textit{i.e.}, $n_{i,\sigma}=\braket{n_{i,\sigma}} + \delta n_{i,\sigma}$, and keeps fluctuations to the first order, giving
\begin{equation}
\begin{aligned}
    \label{HF_Hamiltonian}
    H = &
    -t\sum_{\langle i,j\rangle,\sigma}
    c^\dagger_{i,\sigma}c_{j,\sigma} +\frac{U}{2}\sum_{\substack{{i,} \\{\sigma_1\neq\sigma_2}}}
    (n_{i,\sigma_1} \langle n_{i,\sigma_2}\rangle+\langle n_{i,\sigma_1}\rangle n_{i,\sigma_2})\\
    &-\frac{U}{2}\sum_{\substack{{i,} \\{\sigma_1\neq\sigma_2}}}\langle n_{i,\sigma_1}\rangle \langle n_{i,\sigma_2}\rangle.
\end{aligned}
\end{equation}
Therefore, the problem maps to an effective non-interacting model in the presence of an external field, which can be solved self-consistently, and the ground state is numerically found. 
Periodic boundary conditions are utilized. 
Details of calculations are presented in Appendix~\ref{append:technique}.

\section{Results}
\label{sec:phase_D}
In this section, we present the Hartree mean-field results of unit-density flavor-imbalanced SU(3) Fermi-Hubbard models. 
We investigate flavor density deviations $\delta$ from two configurations: a spin balanced case, so the imbalanced system is characterized by flavor densities $\left(\tfrac{1}{3}-\delta,\tfrac{1}{3}+\delta,\tfrac{1}{3}\right)$, and a case with one component nested, so the flavors are $\left(\tfrac{1}{4}-\delta,\tfrac{1}{4}+\delta,\tfrac{1}{2}\right)$.

\subsection{($\tfrac{1}{3}-\delta,\tfrac{1}{3}+\delta,\tfrac{1}{3}$)}
\label{sec:third}
\subsubsection{Ground state}
\begin{figure}[h]
\includegraphics[width=0.98\linewidth]{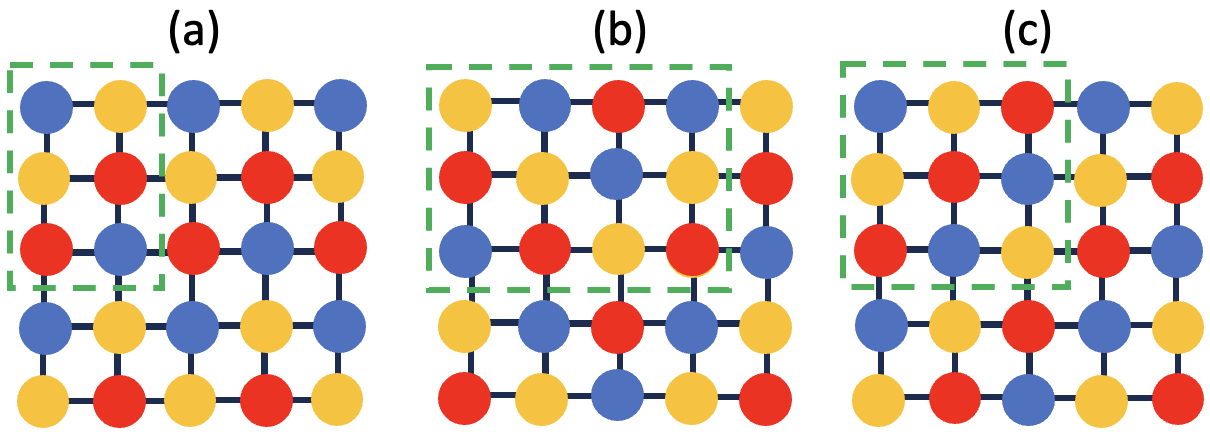}
  \caption{Ground states of the spin flavor balanced SU(3) Hubbard model, where the three colors indicate the flavor that dominates at each site. The three phases are the ground states at: (a) \textit{tooth}, $U\in(3.5t,4.75t)$; (b) \textit{zig-zag}, $U\in(4.75t,5.65t)$; (c) \textit{stripe}, $U>5.65t$. The unit cells of lattice structure are enclosed with dashed lines.  The \textit{zig-zag} phase holds an anisotropic charge density wave, as shown in Appendix~\ref{appx:ground_state_1}. The Hartree calculation is performed on $12\times12$ systems.
  \label{fig:one_third_pattern}}
\end{figure}

Three ordered phases at unit-density are identified at the spin flavor densities of $\left(\tfrac{1}{3},\tfrac{1}{3},\tfrac{1}{3}\right)$, as shown in Fig.~\ref{fig:one_third_pattern} and the same as those found in Ref.~\cite{feng2023metal}, in addition to an unordered \textit{metallic} phase. Transitions among these three phases, named as \textit{tooth}, \textit{zig-zag}, and \textit{stripe} in Fig.~\ref{fig:one_third_pattern}, are observed as $U$ varies.
While the Heisenberg-limit solution \textit{stripe} is seen in various calculations~\cite{toth2010three,sotnikov2015critical,bauer2012three,schlomer2024subdimensional}, two other phases at moderate $U$ (\textit{tooth} and \textit{zig-zag}) were also discovered in Ref.~\cite{feng2023metal}, which are described by $2\times 3$ and $3\times 4$ unit cells, respectively. 
We characterize these phases by the spin flavor structure factor, defined as 
\begin{equation}
    S(\vec{k},\sigma) = \frac{1}{\mathcal{N}(\sigma)}\sum_{i,j}\langle n_{i,\sigma}n_{j,\sigma}\rangle\text{e}^{-i\vec{k}\cdot(\vec{R}_i-\vec{R}_j)},
\end{equation}
where $\vec{k}$ is a wavevector, $\mathcal{N}(\sigma)$ is the total particle number of flavor $\sigma$, and $\vec{R_i}$ is the lattice position of site $i$. The summation of sites $i$ and $j$ is over the cell of calculation.
Both \textit{tooth} and \textit{zig-zag} phases contribute to a $(\tfrac{2\pi}{3},\pi)$ peak, and the \textit{zig-zag} phase uniquely contributes to a $(\tfrac{2\pi}{3},\tfrac{\pi}{2})$ peak. The \textit{stripe} phase contributes to a $(\tfrac{2\pi}{3},\tfrac{2\pi}{3})$ peak. As $\delta$ is increased towards $\tfrac{1}{6}$, the second spin flavor with population $(\tfrac{1}{3}+\delta)$ becomes close to half-filling and gives rise to a phase with a checkerboard pattern in this spin flavor (hereafter named \textit{SU(2)-N\'eel}), which gives a $(\pi,\pi)$ peak in the structure factor.

\begin{figure}
\includegraphics[width=0.97\linewidth]{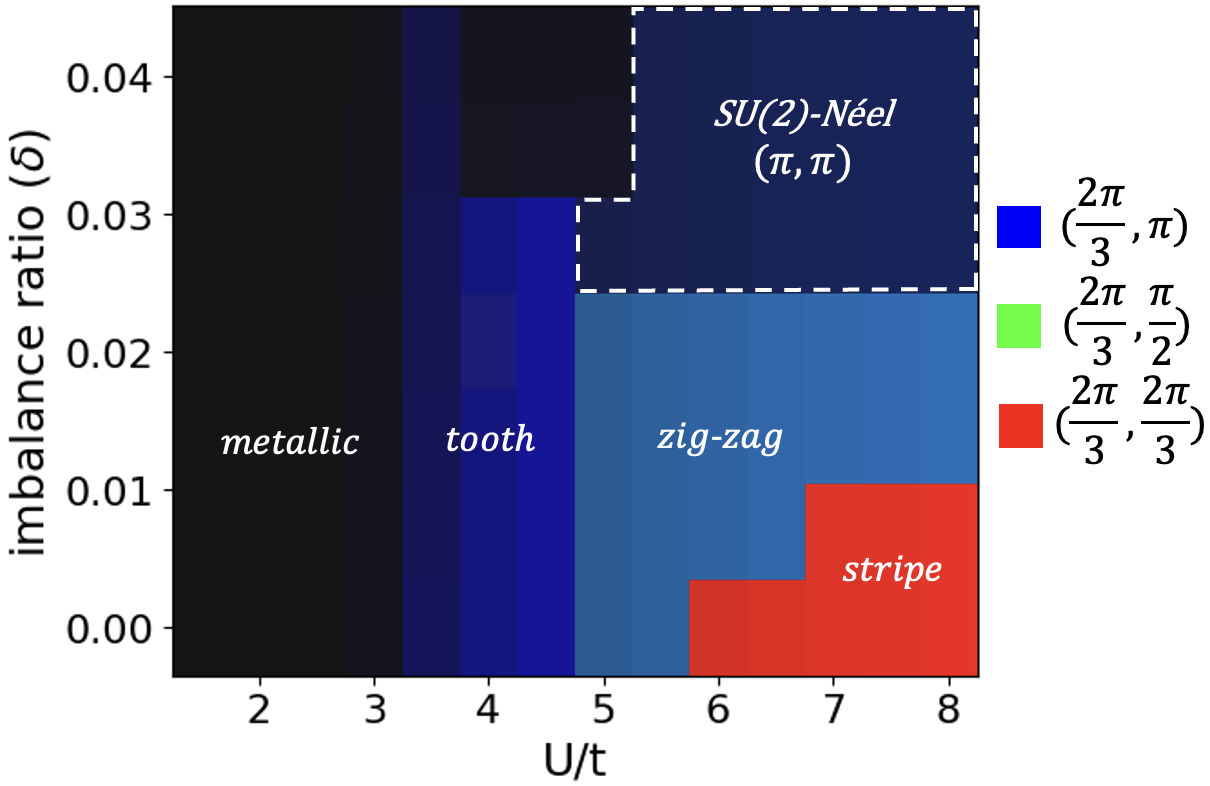}
  \caption{$\left(\tfrac{1}{3}-\delta,\tfrac{1}{3}+\delta,\tfrac{1}{3}\right)$. RGB color gives the value of the structure factor per atom for flavor 3 in the ground state at $\vec{k}=(\tfrac{2\pi}{3},\pi)$, $(\tfrac{2\pi}{3},\tfrac{\pi}{2})$, and $(\tfrac{2\pi}{3},\tfrac{2\pi}{3})$, respectively, normalized by the highest peak (which is the $(\tfrac{2\pi}{3},\tfrac{2\pi}{3})$ peak) found among all $\vec{k}$ values and $(U, \delta)$ combinations. Similar plots with the other two flavors give the same phase diagram. To make the small peaks more visible, the brightness reflects the square root of the peak heights. The \textit{zig-zag} region has both blue and green peaks.  
  The second and third flavors in the region enclosed by white dashed line have significant $(\pi,\pi)$ peaks, which are not shown because of the limit of RGB coloring. Details of the $(\pi,\pi)$ peaks can be found in Appendix~\ref{append:one_third_example}. The size of this Hartree calculation is $12\times 12$.   }
  \label{fig:ground_PD}
\end{figure}

Monitoring these structure factors provides qualitative insights into how the ground states are affected by the imbalance parameter $\delta$. 
The magnitudes of structure factors for the ground states as a function of $U$ and $\delta$ are plotted in Fig.~\ref{fig:ground_PD}.
At small interaction $U<3.50t$, a \textit{metallic} phase is identified. Although other peaks are not shown in Fig.~\ref{fig:ground_PD}, we have verified that the region of small $U$ gives no significant structure factor peaks and thus remains paramagnetic.

\begin{figure}[h]
\includegraphics[width=0.7\linewidth]{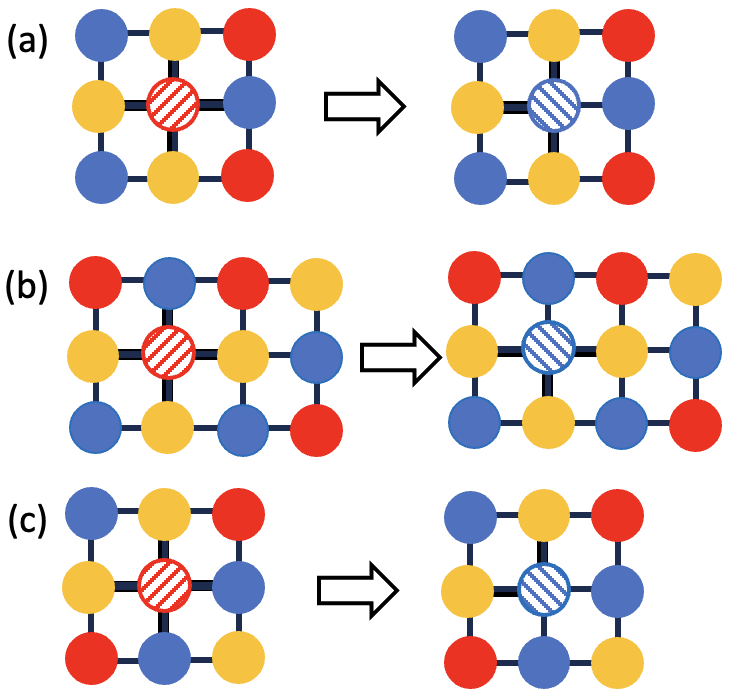}
  \caption{Phase stability against imbalance. (a) \textit{tooth}; (b) \textit{zig-zag}; (c) \textit{stripe}. The diagrams illustrate the replacement of a red-flavor atom with a blue-flavor atom. At relatively large $U$, where the energy is mainly determined by the nearest-neighbor spin correlation, the system can lower its energy through spin-flavor superexchange (on the bold bonds). In both \textit{tooth} and \textit{zig-zag} phases, after introducing the flavor imbalance, the newly added blue atom is surrounded by three neighboring sites occupied by a different (yellow) flavor. In contrast, in the \textit{stripe} phase, the new blue atom has only two distinct neighbors with differing flavors, with which it can undergo superexchange. This demonstrates that the \textit{stripe} phase is more energetically penalized by imbalance than the \textit{tooth} or \textit{zig-zag} phases.
  \label{fig:local_imbalance}}
\end{figure}

All three ordered phases survive at small flavor imbalances: the \textit{tooth} and \textit{zig-zag} phases with $(\pi,\tfrac{2\pi}{3})$ peaks are more stable than the \textit{stripe} phase in the presence of flavor imbalance, as they survive utill $\delta$ goes to approximately 0.022 and beyond. The \textit{stripe} phase with $(\tfrac{2\pi}{3},\tfrac{2\pi}{3})$ peak may be observed at $U\ge7.0t$ for $\delta<0.01$, but for $\delta\in(0.01,0.022)$, the system prefers the \textit{zig-zag} phase to the \textit{stripe} phase. A cut at $U=7.0t$ is shown in Appendix~\ref{append:one_third_example} to demonstrate how the system transitions from the \textit{stripe} phase to the \textit{zig-zag} phase and then builds up $(\pi,\pi)$ peaks.

The qualitative structure of the phase diagram can be understood by considering how imbalance affects the ability of each flavor to delocalize and thereby lower the energy, as illustrated in Fig.~\ref{fig:local_imbalance}. At large $U$ and small $\delta$, imbalance can be viewed as introducing a dilute gas of particles with a differing spin (\textit{i.e.} a particle-hole pair, where a hole is created in the original spin flavor and a particle is created in a new one), and these particles can be treated as independent. When a site’s particle is replaced by another particle with a different spin flavor, the added particle in the \textit{tooth} and \textit{zig-zag} phases can engage in superexchange with three neighboring sites, as restricted by the Pauli exclusion principle, thereby reducing its local energy relative to a localized particle. The mobility is further restricted in the \textit{stripe} phase, where the added particle can hop to only two neighboring sites, leading to a higher local energy than that in the \textit{zig-zag} phase. This explains why increasing $\delta$ will eventually cause the \textit{stripe} phase to be unstable to the \textit{zig-zag} phase. This reasoning also explains the nearly vertical phase boundary between \textit{tooth} and \textit{zig-zag}: both phases respond identically at the local level to small flavor imbalance. Similar superexchange counting arguments determine domain wall structures in bosonic systems~\cite{zhu20self}.

\textit{N\'eel} ordering with $(\pi,\pi)$ peaks, similar to the SU(2) case, is built without exact nesting at $U>5.0t$ and $\delta\gtrsim0.022$. When $\delta$ is sufficiently large, the second flavor with population density $\left(\tfrac{1}{3}+\delta\right)$ approaches half-filling ($\delta=\tfrac{1}{6}$), where the nesting leads a clear transition to a phase with $(\pi,\pi)$ peaks. This shows the breakdown of the local perturbative description of flavor imbalance in Fig.~\ref{fig:local_imbalance} when $\delta$ is not small.

In addition to the spin order, we observe that two of the phases---the \textit{zig-zag} and \textit{SU(2) N{\'e}el} phases---have accompanying charge order. The \textit{zig-zag} charge order is an anisotropic two-sublattice order characterized by a $\vec{k}=(\pi,0)$ ordering wavevector, while the \textit{N{\'e}el} charge order is in a checkerboard pattern characterized by $\vec{k}=(\pi,\pi)$. We present and explain the observed behavior in Appendix~\ref{appx:ground_state_1} by showing symmetry arguments that the $(\pi,0)$ charge order is disallowed in the \textit{tooth} and \textit{stripe} phases, while allowed in the \textit{zig-zag} phase. The $(\pi,\pi)$ charge order gets allowed when the spin \textit{N{\'e}el} order develops with a single flavor near half-filling, and smoothly connects to the order observed in Sec.~\ref{sec:one_half} when one of the spin flavors is exactly half-filled.

\subsubsection{Thermal states}
\label{sec:thermal_phase}
\begin{figure}[h]
\includegraphics[width=0.82\linewidth]{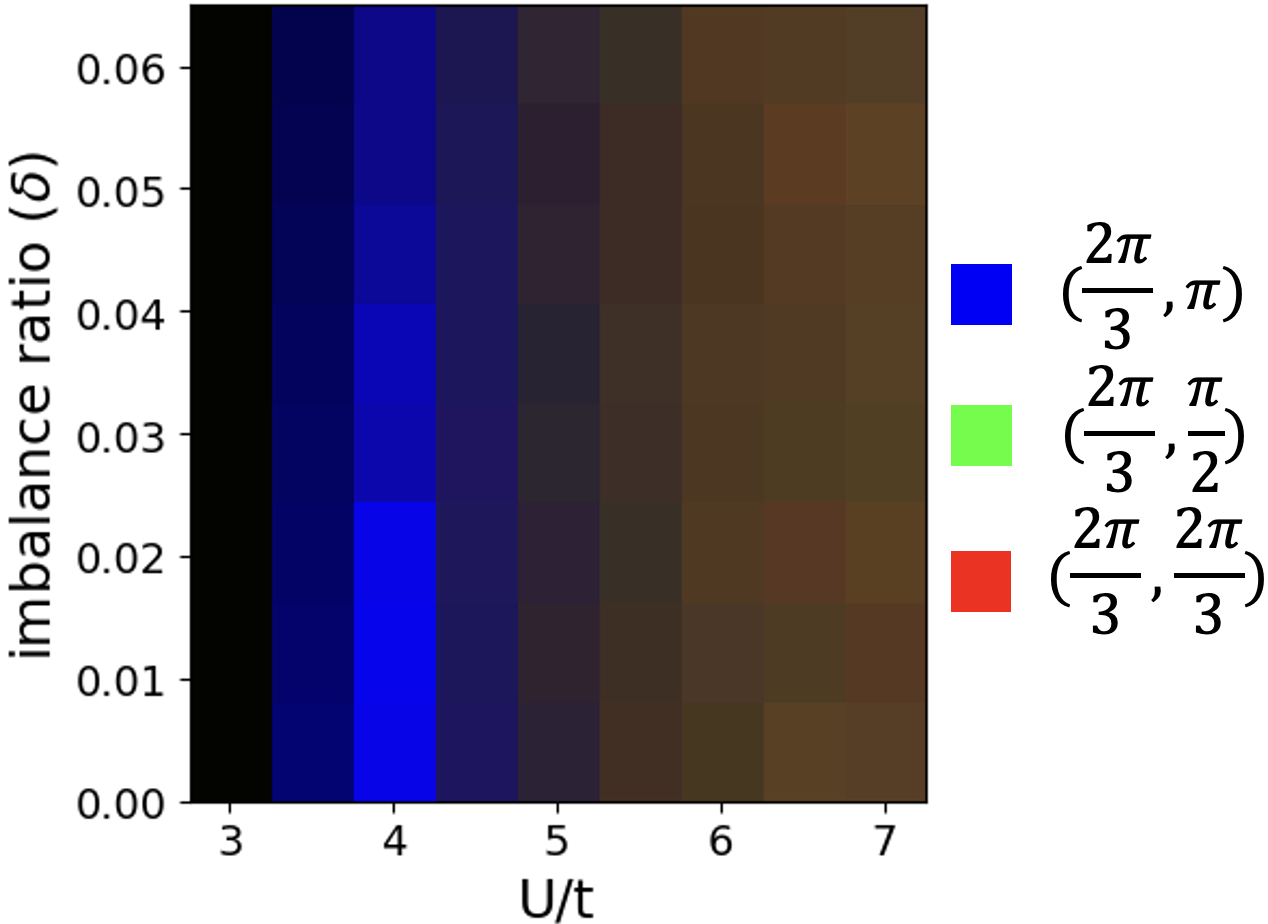}
  \caption{$\left(\tfrac{1}{3}-\delta,\tfrac{1}{3}+\delta,\tfrac{1}{3}\right)$. Thermal-averaged values of the structure factor peaks per atom 
  at $T=0.1t$, normalized by the highest peaks found among all $\vec{k}$ values and $(U, \delta)$ combinations. The plotted structure factor peak height is for the third flavor, but the other two flavors give the same phase diagram. There is a high peak of $(\tfrac{2\pi}{3},\pi)$  dominating in the $U\in(3.5t,4.5t)$ and small $\delta$. 
  The brightness of the green and red peaks are magnified by a factor of 5 for visibility.
  The $(\pi,\pi)$ peaks (not shown) are small compared to the $(\tfrac{2\pi}{3},\pi)$ peak shown in this figure.  
  \label{thermal_PD}}
\end{figure}

The mean-field results may provide insight into the properties of low-energy states in both numerical methods and experiments. For systems with multiple competing low-energy states at similar energy scales, locating a ground state can be challenging in some variational methods and in state-preparation experiments. These methods may eventually get stuck in solutions with distinct local orders in different regions. As such, sampling low-energy states in our calculation gives some indication of possible outcomes. Therefore, a thermal-averaged analysis over multiple randomly generated Hartree solutions can indicate what is likely to be observed experimentally.
We generate at least 100 self-consistent zero-temperature Hartree solutions at each $(U,\delta)$ data point, drawing initial conditions randomly\footnote{For initial conditions of the Hartree calculation, we assign a uniformly-picked random number for each flavor to each site of a $12\times24$ unit cell. See Appendix~\ref{append:technique} for details.}, and then perform a thermal average of these zero-temperature solutions at $T=0.1t$, which corresponds to the lowest temperature where state-of-the-art optical lattice experiments are able to achieve~\cite{taie2022observation}. The results are plotted in Fig.~\ref{thermal_PD}.

The thermal phase diagram shows that, in the moderate interaction strength region ($3.5t\lesssim U\lesssim4.5t$), the $(\tfrac{2\pi}{3},\pi)$ peak is obvious at the temperature of $T=0.1t$. 
This is evidence of the remnant of the \textit{tooth} phase, and the lack of other peaks suggests the difficulty of observing the other two ordered phases. 
Similar to the ground-state results, when $U<3.5t$, no spin order is observed, and the system is in a \textit{metallic} phase.
When $U\gtrsim5.0t$, the other two peaks $(\tfrac{2\pi}{3},\tfrac{2\pi}{3})$ and $(\tfrac{2\pi}{3},\tfrac{\pi}{2})$ can barely be observed, with the highest peaks in both signals are less than 10\% percent of the highest $(\tfrac{2\pi}{3},\pi)$ peak.  
Thus, if ground states are sampled in a way similar to this numerical ensemble, only the \textit{tooth} phase robustly reveals itself in the structure factor at these temperatures, and only in a window of $U$ values. To observe the other two ordered phases, lower temperature is needed. 

The missing signals of the ordered phases at large $U$ are caused by the
fact that the mean-field solutions often get stuck in metastable states with local ordering, 
which is similar to obstacles existing in state-preparation experiments. As evidenced by some examples in Appendix~\ref{append:excitation}, typical excited Hartree solutions are different for moderate interaction strength [roughly $U\in(3.5t,4.5t)$] and large interaction strength. For the former, these excited solutions are mostly the corresponding ground states with structural distortion, such as defects or domain walls, thus keeps part of the structure factor peaks; For the latter, when the tunneling is small, most of the excited solutions are trapped in local \textit{SU(2) N\'eel} order domains with two of the three flavors forming checkerboard patterns, thus lost all SU(3) signals.

The structural distortion in these excited solutions suggests that QGM may offer more reliable detection of potential order in SU($N$) Fermi-Hubbard models than standard structure-factor measurements~\cite{mikkelsen2023relation}. 
When the temperature in AEA optical lattice experiments is not sufficiently low, the structure factors may be weakened by the presence of defects or domain walls. QGM, however, permits single-shot analyses of individual configurations in the real space and 
thus helps to identify the phases more directly.

\subsection{($\tfrac{1}{4}-\delta,\tfrac{1}{4}+\delta,\tfrac{1}{2}$)}
\label{sec:one_half}
In this section, we study another spin flavor densities $(\tfrac{1}{4}-\delta,\tfrac{1}{4}+\delta,\tfrac{1}{2})$, where the third flavor is at half-filling and therefore its Fermi surface is nested. In contrast to the fast convergence to the thermodynamic limit for $(\tfrac{1}{3}-\delta,\tfrac{1}{3}+\delta, \tfrac{1}{3})$ flavor densities in Sec.~\ref{sec:third}, Hartree solutions for this set of flavor densities suffer more significant finite-size effects at moderate $U$, as detailed in Appendix~\ref{append:convergence}. We take the system size of $24\times24$ for calculation, but some key points (\textit{e.g.}, points along phase boundaries) are also checked on  $36\times36$ lattices.

\subsubsection{Ground state}
\label{sec:ground_state_nested}
\begin{figure*}
\includegraphics[width=0.82\textwidth]{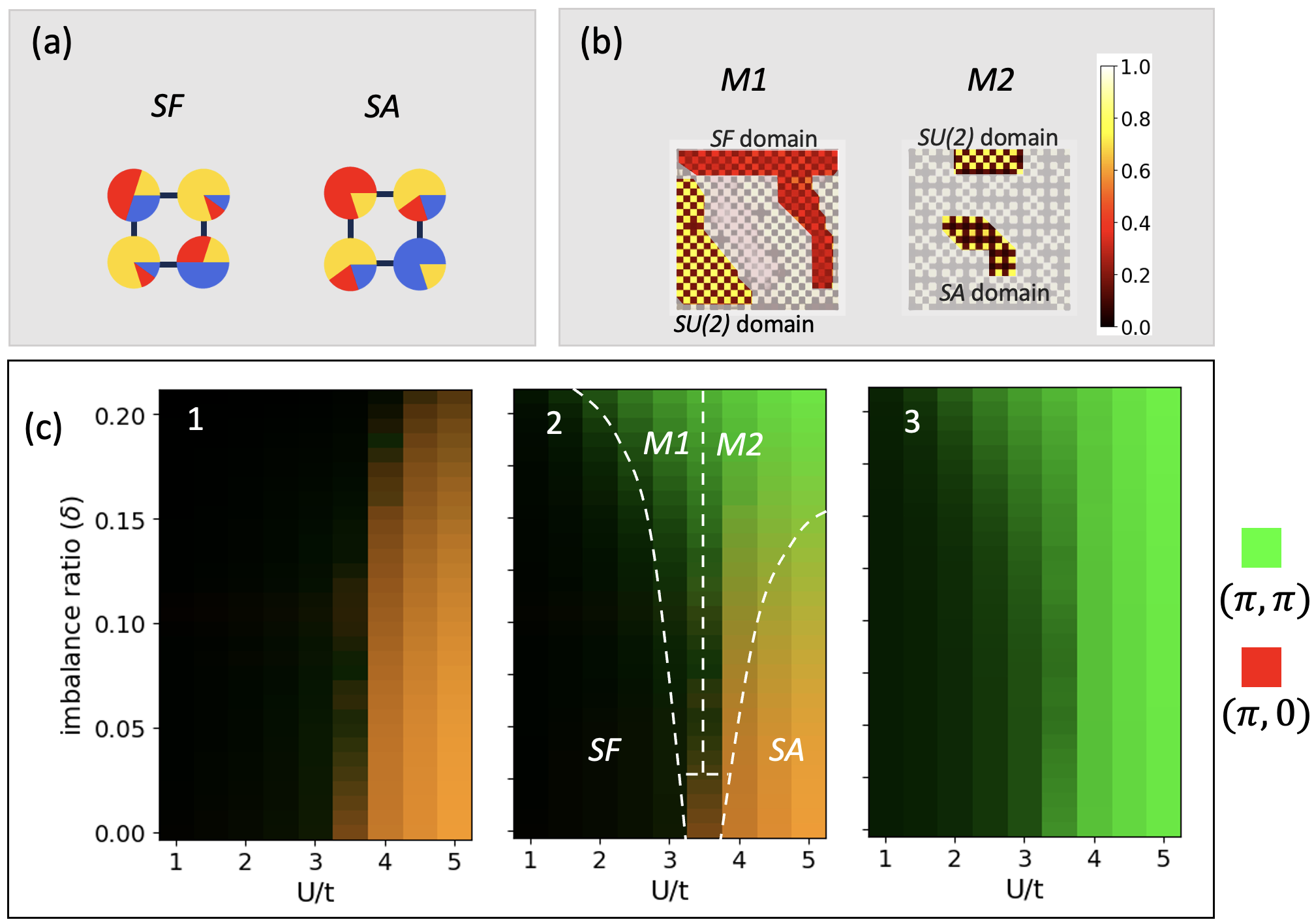}
  \caption{$\left(\tfrac{1}{4}-\delta,\tfrac{1}{4}+\delta,\tfrac{1}{2}\right)$. (a) Ordered phases \textit{SF} and \textit{SA}. (b) Flavor density distributions of example mixtures $M1$ and $M2$. The results are shown for flavor 2, with typical domains highlighted. (c) Structure-factor peaks per atom of the ground states of flavor 1, 2, and 3, respectively, normalized by the highest peak found among all $\vec{k}$ values and $(U, \delta)$ combinations. The orange regions have both peaks. The system size  is $24\times 24$. To make the small peaks more visible, the brightness reflects the square root of the peak heights. The mixture region in the middle has three parts, as specified in the main text.}
  \label{fig:nested_ground}
\end{figure*}

Because the third flavor is half-filled,
it is nested and its density always shows a checkerboard pattern. For the other two flavors (hereafter named ``minor flavors"), two distinct structure-factor peaks [$(\pi,\pi)$ and $(\pi,0)$] are noticeable.  
As evidenced in Fig.~\ref{fig:nested_ground}, the ground states of all flavors manifest finite $(\pi,\pi)$ peaks, although in some small $U$ region these peaks are too small to see. Separately, at $U>3.5t$, a $(\pi,0)$ peak is evident in two minor flavors that exhibit a $2\times 2$ order, but the region of this order shrinks in the phase diagram as the flavor imbalance increases. The orange region in Fig.~\ref{fig:nested_ground} shows both green $(\pi,\pi)$ and red $(\pi,0)$ peaks.

From these two peaks, two ordered phases, named as \textit{superlattice ferromagnetic} (\textit{SF}) and \textit{superlattice anti-ferromagnetic} (\textit{SA}), are identified at different $U$ and $\delta$ settings. As shown in Fig.~\ref{fig:nested_ground}, in the \textit{SF} phase, all three flavors show only the $(\pi,\pi)$ peaks, \textit{i.e.}, the half-filled flavor dominantly occupies one of the two sublattices and the other two evenly occupy the other sublattice. In the \textit{SA} phase, two minor flavors show $(\pi,0)$ peaks, \textit{i.e.}, the half-filled flavor occupies one of the sublattices while the other two flavors alternately occupy the other sublattice. 
The two ordered phases are illustrated in Fig.~\ref{fig:nested_ground}(a). 
In literature studying SU(3) Heisenberg models, the \textit{SF} phase is sometimes referred as ``minority-united canted-N\'eel" (MUCA)~\cite{motegi2023thermal}, which specifies the ferromagnetic order of two minor flavors. Similarly, the pattern in the (\textit{SA}) phase can be understood as antiferromagnetic order of the minor flavors in a diagonal superlattice.
The dependence on $U$ of these two ordered phases suggests that, at small $U$, the system is primarily governed by minimizing the interaction energy between the half-filled flavor and the two minor flavors.
As $U$ grows, this repulsion between the two minor flavors shows up and leads to a superlattice antiferromagnetic order. The discontinuity in Fig.~\ref{fig:nested_balanced} in Appendix~\ref{append:convergence} gives the signal of phase transition between \textit{SF} and \textit{SA} phases.

With increasing $\delta$, the height of the $(\pi,\pi)$ peaks with small $U$ values increases for the second and third spin flavors. Although we cannot calculate for $\delta$ values very close to $0.25$ due to numerical obstacles, 
when $\delta$ approaches $0.25$, the system goes to the unit-density SU(2) limit, consistently resulting in an \textit{SU(2) N\'eel} phase for any finite $U$ values. The connection to SU(2) limit is more clearly discussed with magnetization defined in Sec.~\ref{sec:magn}.

There is a region of mixture in the phase diagram, as indicated in Fig.~\ref{fig:nested_ground}(c). In this region, three phases seem to coexist, including both \textit{SF} and \textit{SA}, as well as the \textit{SU(2) N\'eel} phase. Examples of these phase mixtures have been shown in Fig.~\ref{fig:nested_ground}(b). Qualitatively, this region can be divided into three parts, each of which contains a mixture of two phases, although the precise boundaries are difficult to pinpoint in the phase diagram. For $U$ near $3.5t$ and $0<\delta\lesssim0.03$, a coexistence of \textit{SF} and \textit{SA} phases is observed. 
As $\delta$ increases beyond 0.03, mixtures of \textit{SF} or \textit{SA} phases with the \textit{SU(2) N\'eel} phase appear.  
For approximately $U \lesssim3.5t$, increasing $\delta$ leads to a mixture of \textit{SF} and \textit{SU(2) N\'eel} phases ($M1$). For $U > 3.5t$, increasing $\delta$ instead produces a mixture of \textit{SA} and \textit{SU(2) N\'eel} phases ($M2$). The boundaries separating these mixture regions in the phase diagram are determined 
from real-space images of the ground states, \textit{i.e.}, when multiple imbalance-induced defects cluster together to form an SU(2) domain, the corresponding state is classified as a mixture of the SU(3) and SU(2) phases.

\subsubsection{Magnetism}
\label{sec:magn}
Here we show how the spin correlation in this SU(3) system evolves to the SU(2) limit through the mixture region. Two spin operators are defined as $S_3(i)=n_{i,1}-n_{i,2}$ and $S_8(i)=n_{i,3}-(n_{i,1}+n_{i,2})$ at site $i$, which capture the order for two minor flavors and all flavors, respectively. The subscripts 3 and 8 follow the Gell-Mann convention.

\begin{figure}[h!]
\includegraphics[width=0.99\linewidth]{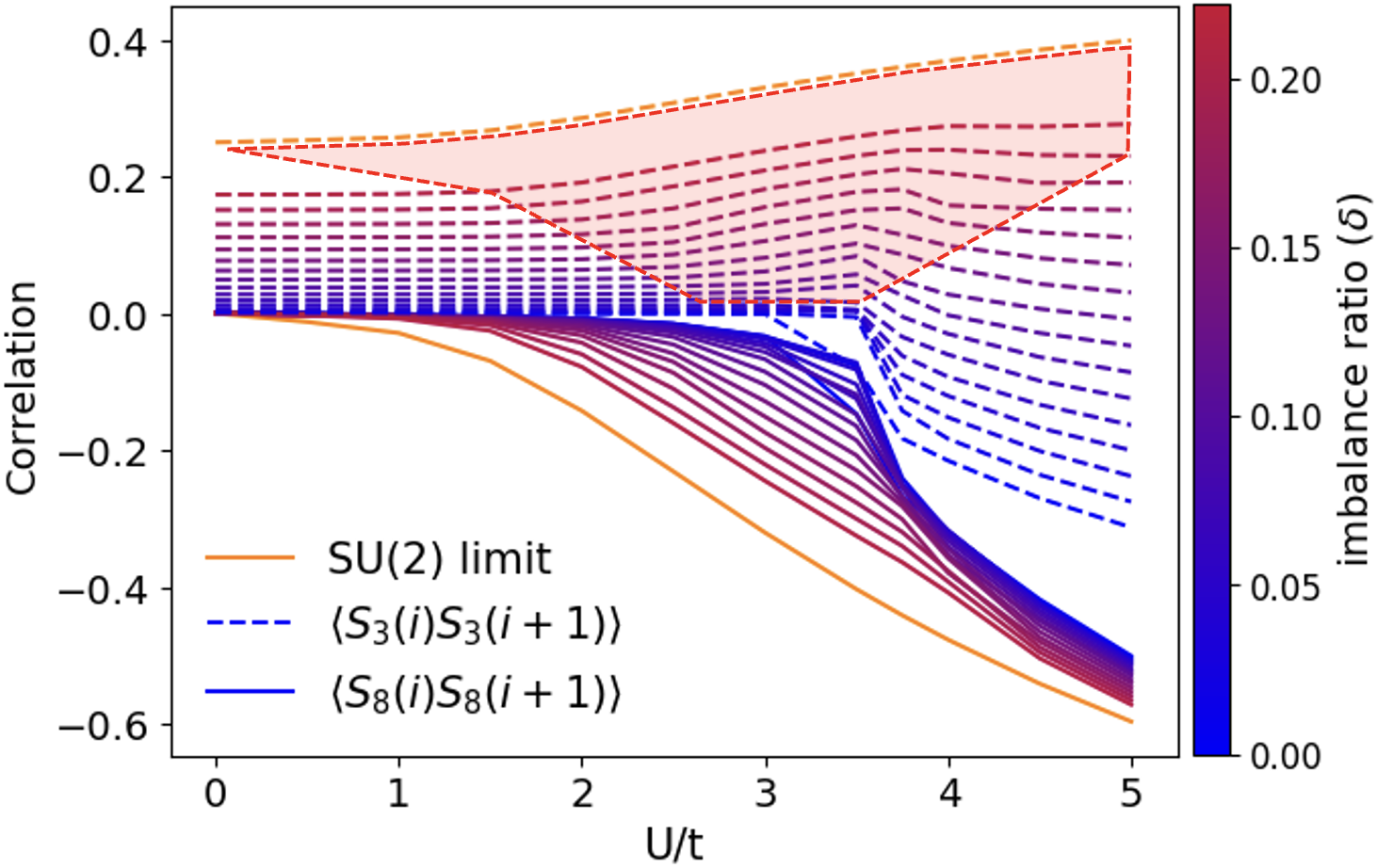}
  \caption{$\left(\tfrac{1}{4}-\delta,\tfrac{1}{4}+\delta,\tfrac{1}{2}\right)$. Nearest-neighbor magnetic correlations $\langle S_m(i)S_m(i+1)\rangle$ for $m=3,8$. The region of mixture is shaded in red according to Fig.~\ref{fig:nested_ground}.
  }
  \label{fig:magetization}
\end{figure}

A connection to the \textit{SU(2) N\'eel} order can be seen in Fig.~\ref{fig:nested_ground}: as the system becomes closer to the half-filled  SU(2) setting, stronger $(\pi,\pi)$ peaks are built in the second and third flavors.
This trend is also clearly captured in the behavior of $S_3$ and $S_8$ correlators, as shown in Fig.~\ref{fig:magetization}.
The negative values of $S_8$ nearest-neighbor correlation indicate some SU(3) ``antiferromagnetic" tendency, which goes to the SU(2) antiferromagnetic order as the $\delta$ value gets to 0.25. On the same set of curves, the jump at small $\delta$ values verifies the transition between \textit{SF} and \textit{SA} phases around $U=3.5t$, as discussed in Sec.~\ref{sec:ground_state_nested}. 
The negative correlations on the $S_3$ curves, an indicator of \textit{SA} phase, are only seen for $U>3.5t$ and $\delta\leq0.1$. In Fig.~\ref{fig:magetization}, the region of mixtures is shaded on the $S_3$ curves according to the observation in Fig.~\ref{fig:nested_ground}.
 
\subsubsection{Thermal states}
\begin{figure}[h!]
\includegraphics[width=0.99\linewidth]{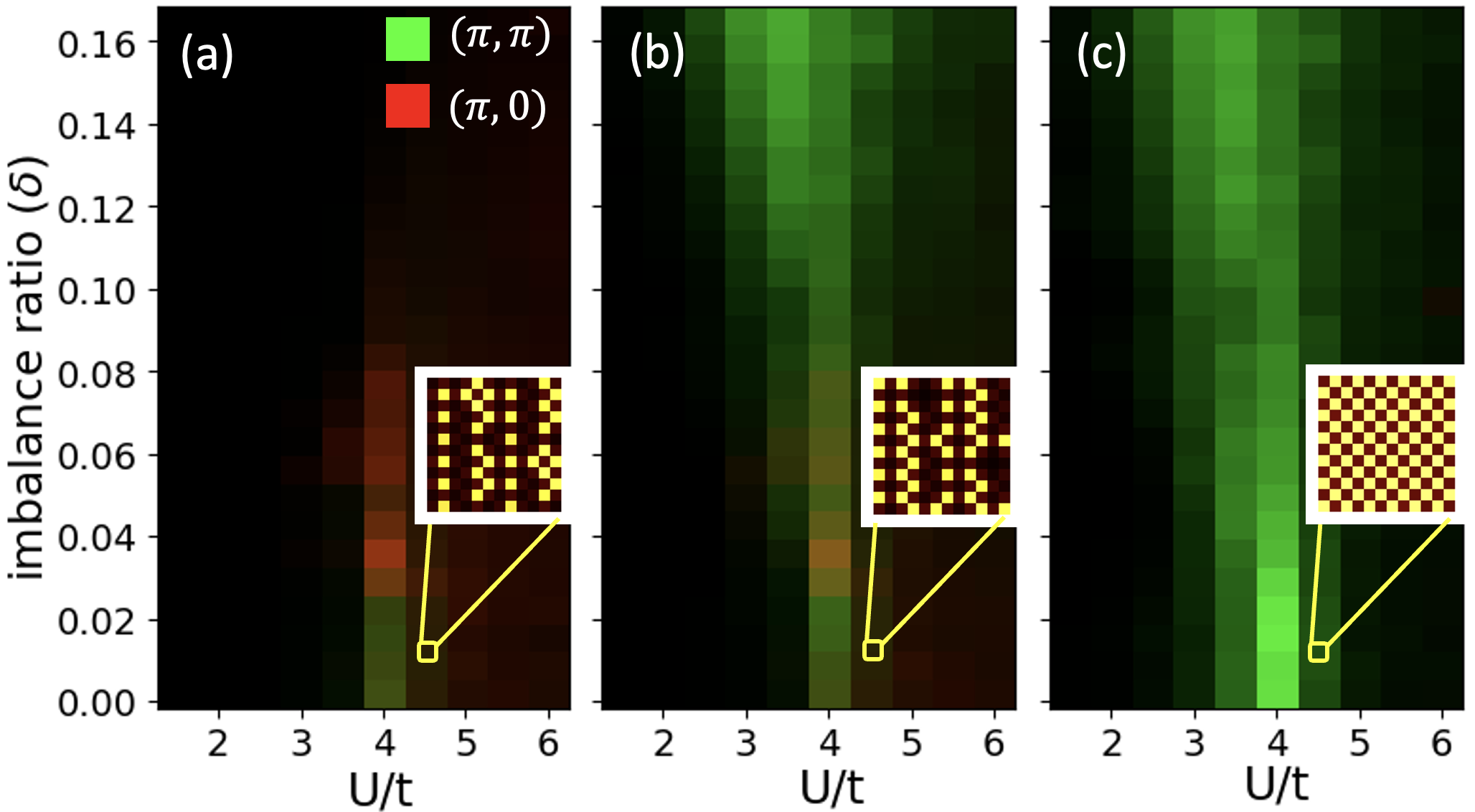}
  \caption{$\left(\tfrac{1}{4}-\delta,\tfrac{1}{4}+\delta,\tfrac{1}{2}\right)$. Thermal-averaged structure-factor peaks per atom at $T=0.1t$, normalized by the highest peak found among all $\vec{k}$ values and $(U, \delta)$ combinations.   (a) Flavor 1, (b) flavor 2, (c) flavor 3. The brightness of the red peaks are magnified by a factor of 5 for clarity. Insets: example of a random Hartree solution with the circled $U$ and $\delta$.
  \label{fig:thermal_nested}}
\end{figure}

The $(\pi,0)$ peak is hard to observe when the system is not cold enough. To see this, we randomly sample at least 100 consistent Hartree solutions at each ($U$,$\delta$) data point and take the thermal average at the temperature of $T = 0.1t$, in the same sense as Sec.~\ref{sec:thermal_phase}. As shown in Fig.~\ref{fig:thermal_nested}, there is only a narrow window for us to see the $SA$ phase with a relatively strong $(\pi,0)$ peak. On the contrary, the $(\pi,\pi)$ peak is clear in the whole moderate $U$ region, and expands as $\delta$ increases. 

Although the $(\pi,0)$ peaks are small in the thermal states, the corresponding \textit{SA} phase remains in some domains, and is directly observable with QGM. 
As the example in the insets of Fig.~\ref{fig:thermal_nested} (also more examples in Appendix~\ref{append:excitation}), 
for $U \ge 3.5t$, these excited states consistently exhibit \textit{SA}-phase domains that can be readily detected in QGM experiments. 
Echoing the discussion in Sec.~\ref{sec:thermal_phase}, this finding highlights the value of QGM in probing ground-state physics in systems such as AEA optical lattices.

\section{Conclusion}
\label{sec:conclusion}
We investigated the unit-density SU(3) Fermi-Hubbard model for two distinct families of spin-flavor densities using Hartree mean-field theory, 
focusing on how spin-flavor imbalance impacts the system's phases. 
We identified three ordered phases when the flavor density is nearly equal ($1/3$ each) 
and two ordered phases when one flavor is nested.

For the densities $\left(\tfrac{1}{3}-\delta , \tfrac{1}{3}+\delta , \tfrac{1}{3}\right)$, 
all three phases remain stable at small flavor imbalance. 
Notably, the \textit{zig-zag} phase demonstrates greater stability to spin-flavor imbalance at large $U$, 
while the \textit{tooth} phase more effectively withstands thermal fluctuations. 
These results highlight the robustness of these newly identified phases, 
even under realistic experimental imperfections. 
Their stability can be understood by treating the imbalanced component as a local perturbation. Checkerboard patterns similar to the \textit{SU(2) N\'eel} phase are also seen when flavor imbalance is large enough.

For the densities $\left(\tfrac{1}{4}-\delta , \tfrac{1}{4}+\delta , \tfrac{1}{2}\right)$, 
the half-filled flavor dominantly occupies one checkerboard sublattice in both \textit{SF} and \textit{SA} phases, 
while the other two minor flavors form different orders---one ferromagnetic and the other antiferromagnetic, on the interleaved sublattice. 
Both phases can be observed when the imbalance is not too large. Also, the phase diagram reveals multiple mixture regions connecting the flavor-imbalanced SU(3) model to the conventional SU(2) framework.

Our thermal ensemble calculations suggest that, at currently accessible temperatures in AEA optical lattice experiments, 
while some ordered phases can still be observed through scattering or other local observable measurements, 
other phases may require QGM to assist detection. 
Looking ahead, alongside ongoing efforts to lower experimental temperatures in AEA optical lattice setups, 
advanced statistical methods~\cite{bohrdt2019classifying,PhysRevA.109.053304} applied to QGM data may offer a powerful route 
to identify new phases.

Our findings offer a roadmap for future experimental efforts aimed at probing these phases. 
With increasing control and cooling available in AEA optical lattices~\cite{muller2021state, yamamoto2024engineering, ahmed2025coherent}, 
the predicted SU($N$) physics may be experimentally studied in the future. 
Beyond these specific platforms, our results could also inspire future investigations 
in other cold-atom settings~\cite{ottenstein2008collisional,mongkolkiattichai2025quantum,mukherjee2024n,mukherjee2024n1} 
and even inform studies in condensed matter systems~\cite{kugel1982jahn,georges2013strong,chen2024multiflavor}. 
The results can also assist sign-problem-free quantum Monte Carlo methods by providing useful trial wavefunctions~\cite{feng2023metal}.
Moreover, the local perturbative picture for flavor imbalance may give insight into the structure of the spin-balanced phase diagram when including longer-distance superexchange, and perhaps help to elucidate the effect of doping in relevant systems~\cite{schlomer2024subdimensional,koepsell2021microscopic,bourgund2025formation}, as replacing a hole in SU(2) systems with a third-flavor particle helps to separate hopping and interaction terms. The results, especially the robustness of phases, may also help to understand the impact of flavor imbalance on the thermalization of difference phases~\cite{huang2020suppression}. We may also consider extending the calculation to other lattice geometries~\cite{rapp2011ground,bauer2012three,botzung2024exact,weichselbaum2018unified} featuring flavor-imbalanced physics or typical larger $N$ values~\cite{corboz2011simultaneous,unukovych20214,weichselbaum2018unified,wang2019slater} and non-trivial gauge fields~\cite{chen:synthetic_2016,zhou2018mott,mamaev2022resonant}.

\newpage

\begin{acknowledgments}
K.R.A.H. and Z.Z. acknowledge support from the National Science Foundation (PHY-1848304),  and the W. M. Keck Foundation
(Grant No. 995764). K.R.A.H. benefited from discussions at the the Aspen Center for Physics, which is supported in part by the National Science Foundation (PHY-1066293). R.T.S. is supported by the grant DOE DE-SC0014671 funded by
the U.S. Department of Energy, Office of Science. Z.Z. thanks Henning Schl\"{o}mer for conversations about calculation and Chunhan Feng for cross-checking balanced mean-field results.
\end{acknowledgments}

\appendix
\section{Convergence criteria and acceleration}
\label{append:technique}
We run Hartree calculations at fixed particle numbers for each flavor. 
The procedure is initialized using a chosen site- and flavor-dependent density configuration $n_{i,\sigma}(\text{iter}=0)$, 
which either contains repeating unit-cell patterns (see Appendix~\ref{append:sublattice}) for the ground-state calculation 
or is uniformly randomized for the thermal phase diagram. 
Given a density configuration $\langle n_{i,\sigma}(\text{iter}=p)\rangle$ in the $p^\text{th}$ iteration, we compute the wavefunctions of 
the Hartree Hamiltonian in Eq.~\eqref{HF_Hamiltonian}, generate a new density configuration according to Fermi-Dirac statistics~\cite{PhysRevB.108.035139}, 
and then feed back the density configuration into the Hartree calculation. 
This self-consistent iteration continues until convergence at the $(p+d)$ round, defined by 
\begin{equation}
\sum_{i,\sigma} \Bigl\lvert \langle n_{i,\sigma}(\text{iter}=p)\rangle 
  - \langle n_{i,\sigma}(\text{iter}=p+d)\rangle\Bigr\rvert < \Delta.
\end{equation}
For a $12\times 12$ lattice, we typically choose $\Delta = 10^{-7}$ and $d = 5$.

\subsection{Anderson acceleration}
To accelerate the rate of convergence, we apply the Anderson acceleration technique~\cite{anderson1965iterative} to the iterations. Instead of using the density configuration  $\langle n^0_{i,\sigma}(\text{iter}=p)\rangle$, which is derived from iteration $p$ to calculate the effective potentials in Eq.~\eqref{HF_Hamiltonian} in step $(p+1)$, we use a linear combination of two sequential iterations as follows,
\begin{align}
&\langle n_{i,\sigma}(\text{iter}=p)\rangle\nonumber\\
=& (1-\alpha)\langle n^0_{i,\sigma}(\text{iter}=p)\rangle+\alpha \langle n_{i,\sigma}(\text{iter}=p-1)\rangle,
\end{align}
and use the result $\langle n_{i,\sigma}(\text{iter}=p)\rangle$ for the next iteration.
Besides accelerating convergence for small $U$, the Anderson method helps to overcome a practical issue, which is often seen at large $U$, that under naive iteration the mean-fields oscillate between two configurations without settling into a steady-state. We empirically find that an efficient $\alpha$ for this calculation is around 0.35. 

\subsection{Initial iteration seeds}
\label{append:sublattice}
To search for Hartree ground states with possible symmetries, we impose repeating blocks of $l\times m$ size to the initial  density configuration $\langle n_{i,\sigma}(\text{iter}=0)\rangle$ with all reasonable combinations of integers $l$ and $m$. We do hundreds of trials for small $U$ (typically $U<4t$) and thousands of trials for large $U$ where, in each trial, the density in the repeating block is uniformly randomly generated. Also, to enable possible solutions without perfect lattice structures, especially when the system is flavor-imbalanced, small perturbations are added to initial states after imposing repeating block structures: $n'_{i,\sigma}(\text{iter}=0) = n_{i,\sigma}(\text{iter}=0)\times(1+r_{i,\sigma})$, where $r_{i,\sigma}$ follows a uniform distribution between $\pm0.05$. 
For each scenario, the lowest-energy Hartree solution is picked as the mean-field ground state.

\section{Ground states}
\label{appx:ground_state}
\subsection{$\left(\tfrac{1}{3},\tfrac{1}{3},\tfrac{1}{3}\right)$ (balanced)}
\label{appx:ground_state_1}
\begin{figure}[h]
\includegraphics[width=0.96\linewidth]{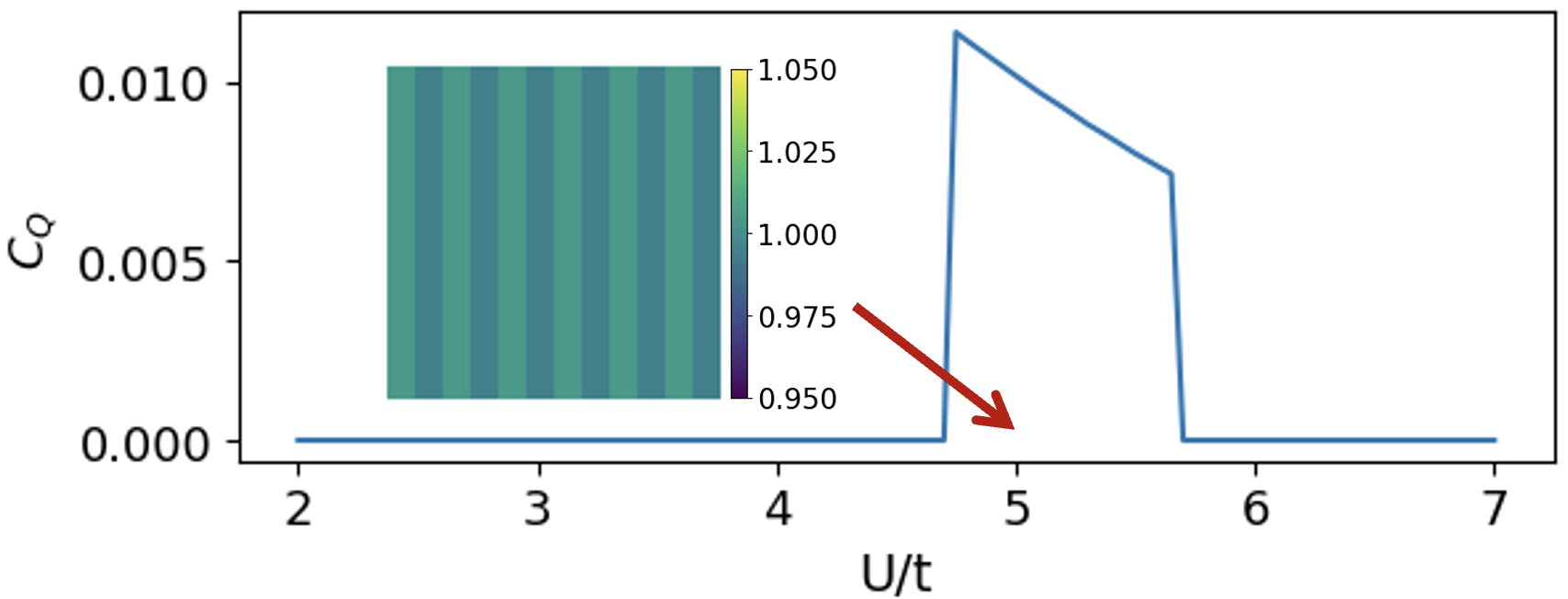}
  \caption{Charge correlation for $(\tfrac{1}{3},\tfrac{1}{3},\tfrac{1}{3})$. The inset gives a typical charge distribution of a flavor-balanced \textit{zig-zag} state at $U=5.0t$.
  \label{fig:one_third_charge}}
\end{figure}

The \textit{zig-zag} state shows anisotropic charge order, different from other flavor-balanced states (\textit{metallic}, \textit{tooth}, and \textit{stripe}). The charge at each site $i$ is summed from all three flavors: $Q_i=\sum_\sigma \langle n_{i,\sigma}\rangle$. To investigate the charge order, we calculate the on-site connected charge correlation 
\begin{equation}
    {\mathcal C}_Q = \frac{1}{L^2}\sum_i\left|Q_i^2-1\right|.
\end{equation}
The results are shown in Fig.~\ref{fig:one_third_charge}.

\begin{figure}[h]
\includegraphics[width=0.9\linewidth]{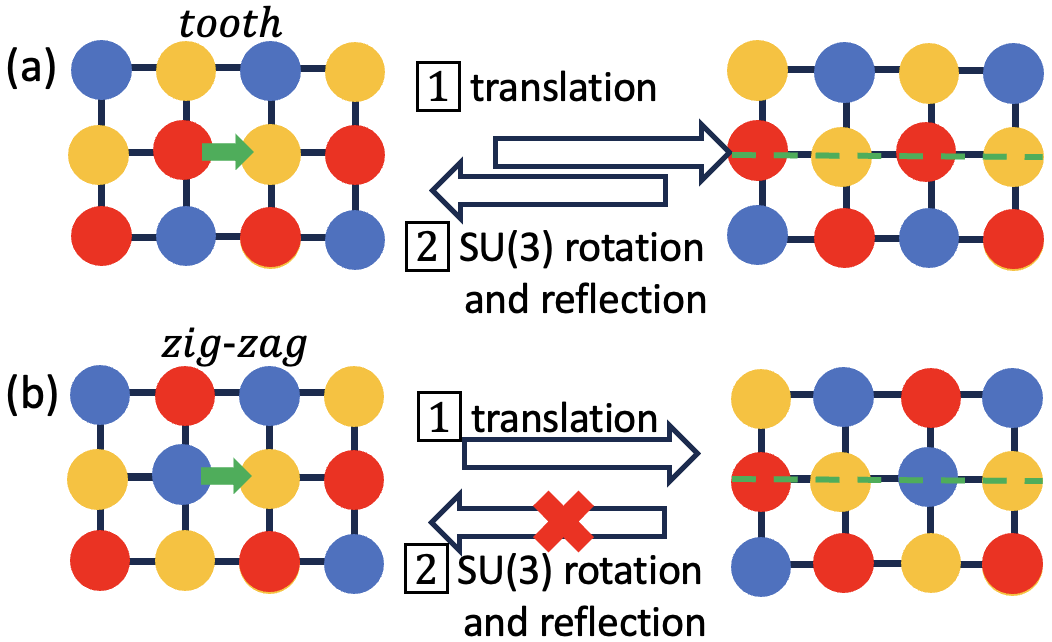}
  \caption{(a) After one-site translation in the horizontal direction, a \textit{tooth} state can be restored to its original structure with an SU(3) rotation and a reflection operation along the horizontal direction. (b) After one-site translation in the horizontal direction, a \textit{zig-zag} state cannot be restored to its original structure with any SU(3) rotation and reflection along the horizontal direction.
  \label{fig:cdw_symmetry}}
\end{figure}

This $(\pi,0)$ CDW order is not allowed in the other two ordered phases (without further spontaneous symmetry breaking) due to their symmetry in the spin orders. For both \textit{tooth} and \textit{stripe} phases, after a translation of one lattice site in either $x$ or $y$ directions, the states can be restored to their original structures by SU(3) rotations and spatial reflection operations along the direction of the translation. This symmetry rules out the existence of any $(\pi,0)$ charge order, in the absence of any further spontaneous symmetry breaking. But the \textit{zig-zag} state does not have such symmetry in one direction and, therefore, allows anisotropic charge order without any additional spontaneous symmetry breaking; put another way, once the \textit{zig-zag} order forms, there is no symmetry reason the anisotropic charge order should vanish, and conversely if charge order formed, it would necessarily perturb the structure of the magnetic order. Two examples are shown in Fig.~\ref{fig:cdw_symmetry}, while the results extend to other states and translation directions.

\subsection{$\left(\tfrac{1}{3}-\delta,\tfrac{1}{3}+\delta,\tfrac{1}{3}\right)$}
\label{append:one_third_example}
\begin{figure}[h]
\includegraphics[width=0.8\linewidth]{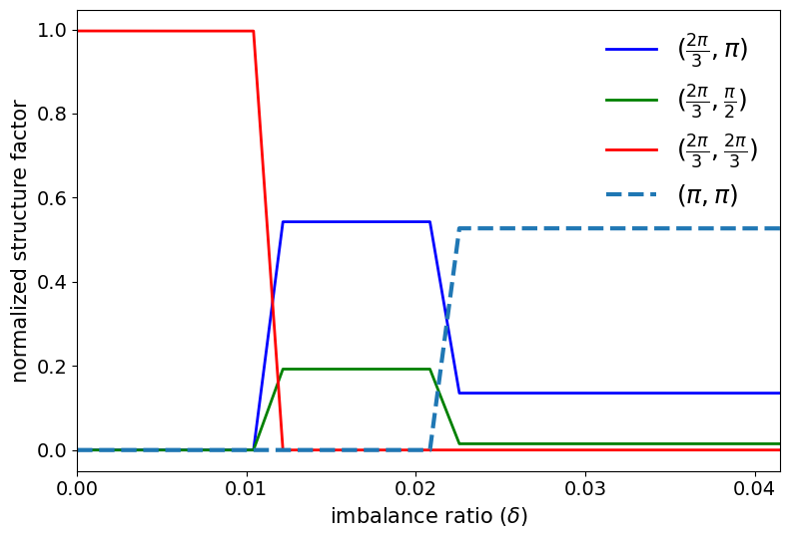}
  \caption{$\left(\tfrac{1}{3}-\delta,\tfrac{1}{3}+\delta,\tfrac{1}{3}\right)$ at $U=7.0t$. A $\left(\frac{2\pi}{3},\frac{2\pi}{3}\right)$ peak indicates the \textit{stripe} phase, the coexistence of $\left(\frac{2\pi}{3},\frac{\pi}{2}\right)$ and $\left(\frac{2\pi}{3},\pi\right)$ peaks marks the \textit{zig-zag} phase, and the $\left(\pi,\pi\right)$ peak gives the checkerboard pattern in density distribution. 
   \label{fig:one_third_U7}}
\end{figure}

Here we show a cut at $U=7.0t$ for the results in Fig.~\ref{fig:ground_PD} in the main text. A peak of $(\pi,\pi)$, which is not specifically shown in the main text due to the limit of RGB coloring, is seen in Fig.~\ref{fig:one_third_U7} when $\delta$ goes beyond 0.022.

\begin{figure}[h]
\includegraphics[width=0.9\linewidth]{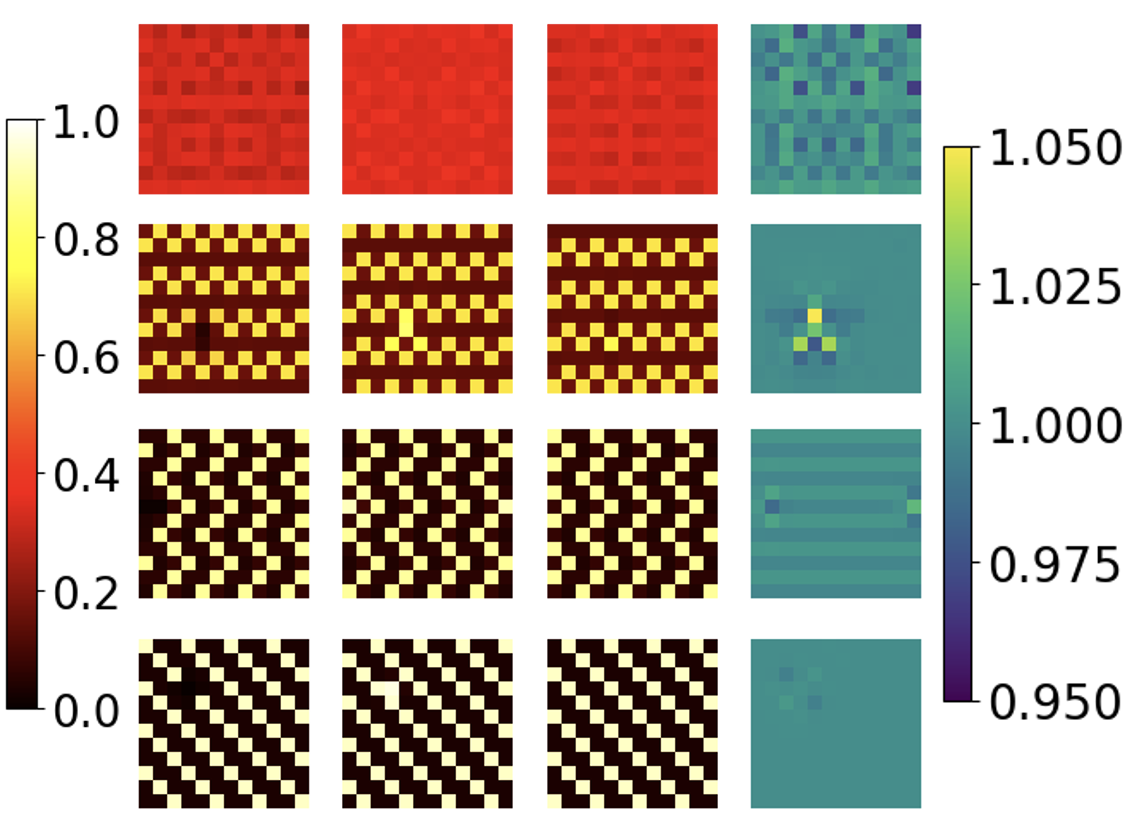}
  \caption{$\left(\tfrac{1}{3}-\delta,\tfrac{1}{3}+\delta,\tfrac{1}{3}\right)$.  Density profiles of ground states when $\delta=\tfrac{1}{144}$, \textit{i.e.}, one-particle imbalance in $12\times12$ systems. From top to bottom: $U=2t$ (\textit{metallic}), $4t$ (\textit{tooth}), $6t$ (\textit{zig-zag}), and $8t$ (\textit{stripe}). Left three: flavor 1-3; right: charge. 
   \label{fig:one_third_example}}
\end{figure}

Examples of typical ground states with nonzero $\delta$ found through the Hartree calculation are shown in Fig.~\ref{fig:one_third_example}. The \textit{metallic} phase keeps its unordered structure at small flavor imbalance. In all three ordered phases, the small flavor imbalance introduces local disorders to the spin and charge structures, which supports the perturbative picture of flavor imbalance, as discussed in Sec.~\ref{sec:third}.

\subsection{$\left(\tfrac{1}{4}-\delta,\tfrac{1}{4}+\delta,\tfrac{1}{2}\right)$}
\begin{figure}[h]
\includegraphics[width=0.95\linewidth]{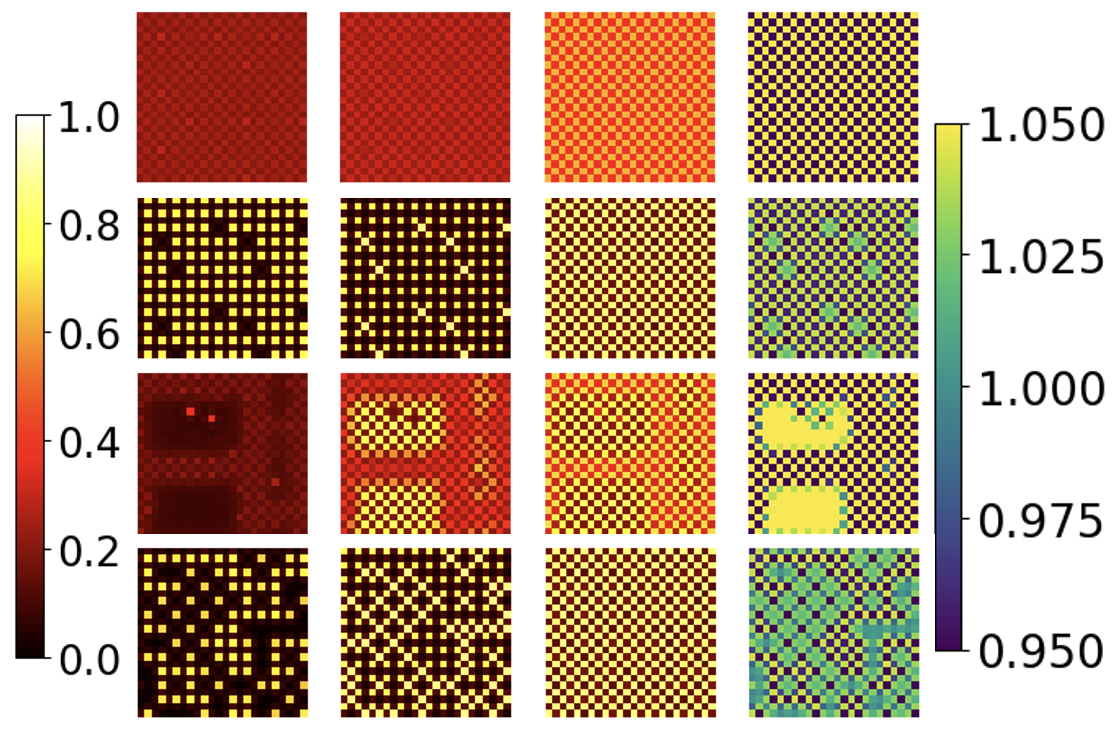}
  \caption{$\left(\tfrac{1}{4}-\delta,\tfrac{1}{4}+\delta,\tfrac{1}{2}\right)$. Density profiles of ground states. From top to bottom: (phase \textit{SF}, $U=3.0t$, $\delta=\tfrac{1}{48}$), (phase \textit{SA}, $U=4.5t$, $\delta=\tfrac{1}{48}$), (mixture $M1$, $U=3.5t$, $\delta=\tfrac{7}{72}$), (mixture $M2$, $U=4.5t$, $\delta=\tfrac{1}{9}$). Left three: flavor 1-3; right: charge. 
   \label{fig:one_nested}}
\end{figure}

Examples of typical ground states of different $U$  and $\delta$ found through the Hartree calculation are summarized in Fig.~\ref{fig:one_nested}. Some solutions are likely to be finite-size artifacts generated by commensurability, as reported similarly in some Hartree-Fock calculation of SU(2) Hubbard models~\cite{PhysRevB.108.035139}.

From Fig.~\ref{fig:one_nested}, we notice that the \textit{SF} and \textit{SA} phases survive in the presence of small flavor imbalances at small ($U\leq3.5t$) and large ($U>3.5t$) interactions, respectively. Increasing $\delta$ gives mixtures of $M1$ and $M2$ at small and large interactions, respectively.

\section{Excited states}
\label{append:excitation}

\begin{figure}[h]
\includegraphics[width=0.99\linewidth]{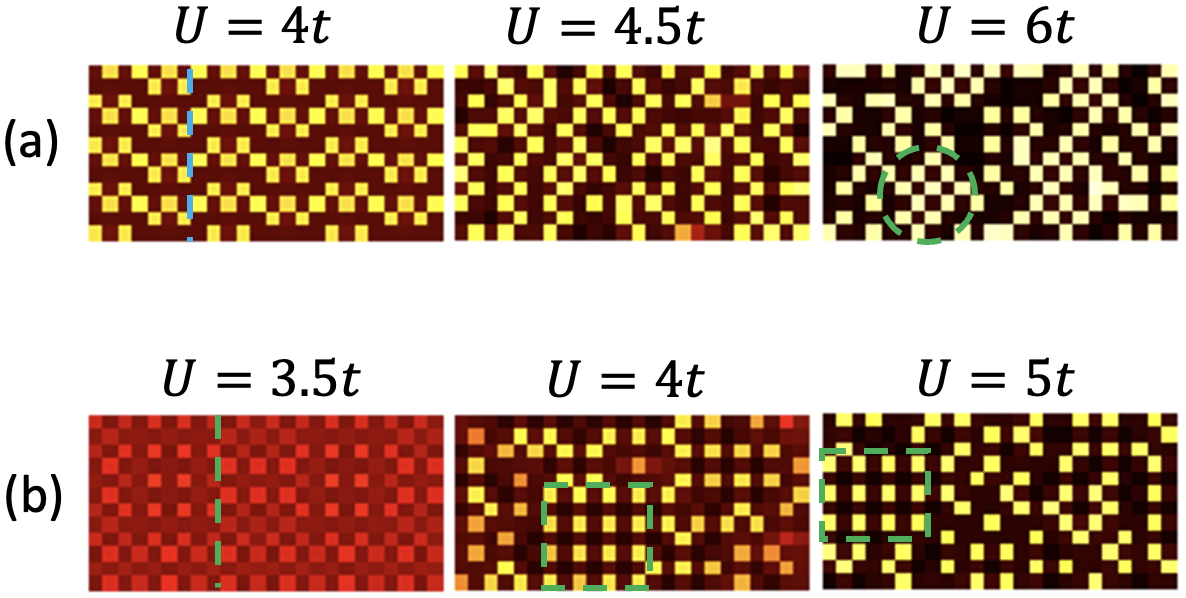}
  \caption{Typical excited Hartree solutions for (a) $(\tfrac{1}{3},\tfrac{1}{3},\tfrac{1}{3})$ and (b) $(\tfrac{1}{4},\tfrac{1}{4},\tfrac{1}{2})$. Results are derived from random seeds, as specified in the main text. Three examples of the first flavor distribution are shown for different $U$. Dashed lines help to locate typical domain walls and ordered domains.
  \label{fig:excited_states}}
\end{figure}

Figure~\ref{fig:excited_states} presents examples of excited Hartree solutions obtained from random initial seeds. 
At $(\tfrac{1}{3}, \tfrac{1}{3}, \tfrac{1}{3})$, \textit{tooth}-phase domains appear for $U=4t$ and $U=4.5t$, 
while the $(\tfrac{1}{4}, \tfrac{1}{4}, \tfrac{1}{2})$ densities consistently display \textit{SA}-phase domains in all three cases. 
The domain walls and disordered regions observed in these solutions diminish the structure factors, 
but QGM can still identify these domains on a shot-by-shot basis, 
facilitating an effective extraction of the ground-state phases at finite temperature or imperfect ground-state preparation.

\section{Convergence}
\label{append:convergence}
\begin{figure}
\includegraphics[width=0.85\linewidth]{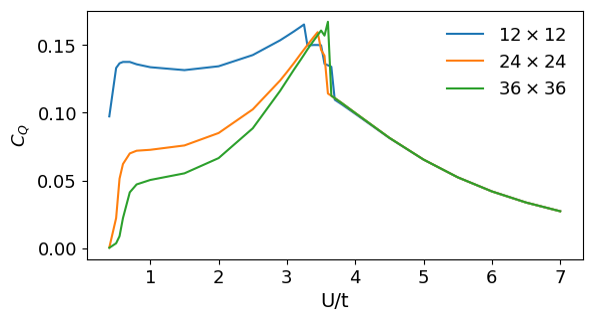}
  \caption{$\left(\tfrac{1}{4},\tfrac{1}{4},\tfrac{1}{2}\right)$. 
  The contrast ${\mathcal C}_Q$ of the ground states. 
  }
  \label{fig:nested_balanced}
\end{figure}

For the densities $\left(\tfrac{1}{4}, \tfrac{1}{4}, \tfrac{1}{2}\right)$, the \textit{SF} and \textit{SA} phases are separated by a phase transition. The on-site connected charge correlation ${\mathcal C}_Q$,
which in this setting reflects the contrast of charge distribution, gives the sign of phase transition between \textit{SF} and \textit{SA} phases. As shown in Fig.~\ref{fig:nested_balanced}, there is a discontinuity on the contrast curves at $U=3.5t$.

The results indicate that the numerics are well converged at large $U$. For small $U$, typically below the phase transition point, the Hartree calculation suffers from noticeable finite-size effects. However, the results give no qualitative differences between $24\times 24$ and $36\times 36$ unit cells.

\bibliography{apssamp}

\end{document}